\documentclass[onecolumn]{IEEEtran}
\usepackage{amssymb}
\usepackage{amsmath,amssymb}
\usepackage{latexsym}
\usepackage{graphicx,subfigure}
\usepackage[dvips]{epsfig}
\usepackage{cite}
\usepackage{epstopdf}
\usepackage{color}
\usepackage{bm}
\usepackage[center]{caption}

\newcommand{\be}{\begin{equation}}
\newcommand{\eq}{\end{equation}}
\newcommand{\ba}{\begin{array}}
\newcommand{\ea}{\end{array}}
\newcommand{\bean}{\begin{eqnarray*}}
\newcommand{\eean}{\end{eqnarray*}}
\newcommand{\bea}{\begin{eqnarray}}
\newcommand{\eea}{\end{eqnarray}}
\newcommand{\nn}{\nonumber}

\newcommand{\R}{\rm {I\kern-2pt R}}

\newcommand{\beq}{\begin{equation}}
\newcommand{\eeq}{\end{equation}}

\newtheorem{theorem}{\bf Theorem}
\newtheorem{remark}{\bf Remark}
\newtheorem{lemma}{\bf Lemma}

\newtheorem{assumption}{\bf Assumption}

\newtheorem{definition}{\bf Definition}

\makeatletter
\newcommand{\Rmnum}[1]{\expandafter\@slowromancap\romannumeral #1@}
\makeatother

\begin{document}
\title{
Distributed Finite-time Differentiator for  Multi-agent Systems Under Directed Graph
}

\author {  Weile Chen,  Haibo Du and Shihua Li
\thanks{ W. Chen is with the School of Electrical Engineering and Automation, Hefei University of Technology, Hefei, Anhui, 230009, and also with the School of Automation, Southeast University, Nanjing, 210096, China.}
\thanks{
H. Du is with the School of Electrical Engineering and Automation, Hefei University of Technology, Hefei, Anhui, 230009, China.}
\thanks{ S. Li is with the School of Automation, Southeast University, Nanjing, 210096, China.}
\thanks{$^*$ Corresponding author: Haibo Du.  {\sl E-mail address: haibo.du@hfut.edu.cn}}
\thanks{$^*$ This work is supported in part by the National Natural Science Foundation of China under Grant Nos. 62073113, 62003122, 62025302 and Natural Science Foundation of Anhui Province of China under Grant Nos. 2008085UD03.}
}
{}

\maketitle

\begin{abstract}
This paper proposes a new {\em distributed finite-time differentiator} (DFD) for multi-agent systems (MAS) under directed graph, which extends the differentiator algorithm from  the centralized case to the distributed case by only using relative/absolute  position information. By skillfully constructing a  Lyapunov function, the  finite-time stability of the closed-loop system under DFD is  proved.
Inspired by the duality principle of control theory,  a distributed  continuous finite-time output consensus algorithm extended from DFD for  a class of  leader-follower  MAS is provided, which not only completely suppresses  disturbance, but also avoids chattering. Finally,  several simulation examples are given to verify the effectiveness of the DFD.

\end{abstract}


\vspace{1.5ex}
\section{Introduction}

The differentiator  means that for a real-time measurable signal $f(t)$, design an algorithm to estimate
$\dot f(t)$ under certain conditions.
Based on the second-order sliding mode algorithm, namely the super-twisting algorithm \cite{Levant_1993}, a famous differentiator algorithm was proposed in \cite{Levant_1998}.
Considering that many mechanical systems can be modeled by a  second-order system,
 the super-twisting algorithm was employed to solve the  control problem in  \cite{davila_tac_2005}.
For the high-order systems, the corresponding differentiator algorithm and output feedback
 control algorithm  were introduced in  \cite{Levant_2003}. In order to accelerate the convergence speed, uniformly convergent differentiators were proposed in \cite{Moreno-Fridman_tac-2011,Moreno-Fridman_auto-2013}.

The above differentiator can be regarded as  centralized differentiator. With the development of science and technology,  networks are playing an increasingly important role \cite{Pei-SCTS-2021}.
In the early stage, in   \cite{Hong-auto-2006, Hong-auto-2008},  the asymptotical  consensus  of second-order leader-follower MAS was realized by designing distributed observers.
The distributed observers for linear systems were studied in \cite{Pei-SCTS-2021,Park-tac-2017,Mitra-tac-2018}.
In   \cite{Liu-TCYB-2018}, the cooperative output regulation of LTI plant was solved   based on the distributed observer.
Considering the uncertainty and disturbance of the system, robust distributed observers were  proposed in
\cite{Hong-TSMC-2022,Wang-tac-2022}.
Combining with adaptive control method,  adaptive distributed observers were proposed in \cite{He-Auto-2017,Lv-TCSI-2020,He-auto-2022}.
In addition, distributed finite-time and fixed-time observers were proposed in \cite{Silm-Tac-2019,Du-Wen-auto-2020}, respectively. However, most of these observers need the leader's internal state information (equivalent to $\dot f(t)$) or control input information (equivalent to $\ddot f(t)$), so they are not distributed differentiators. To the best of our knowledge, no related algorithm can achieve the same function, i.e., {\em distributed finite-time differentiator}.

In the cooperative control of  leader-follower MAS, the full states or partial internal states of leader  are required in most works, which precludes many practical applications where only the output of the leader system is available \cite{Cai-Huang-auto-2021}. In practice, sometimes only the relative position information can be obtained rather than  the absolute global position, and for each follower agent, the more important information is the relative position and relative velocity between itself and the leader, rather than the absolute global position and absolute global velocity. For example, for groups of mobile robots,  the  global positions of the robots are usually not available while  the relative position measurements should be used  instead \cite{Liu-Jiang-auto-2013}. In addition, it is more difficult to get velocity and acceleration measurements  than position measurement \cite{Ren-tac-2012,Ma-Xu-auto-2023} and the follower agents might not be equipped with  velocity sensors to save space, cost and weight \cite{Mei-Ren-Ma-auto-2013,Mei-Ren-auto-2013}.
Therefore, the study of distributed observer and controller based only on the relative position information has important theoretical significance and practical value \cite{Lv-Wen-auto-2020}.

The main  contributions of this paper are given as follows.
Firstly,   unlike the centralized finite-time  differentiator, a framework of {\em  distributed finite-time differentiator} (DFD) is proposed, which can achieve  the exact differential estimation only if  the differentiable signal $f(t)$ is available for at least one agent.
   The  distributed finite-time differentiator can be realized via  the  absolute position  information
    or relative position information, which is more available for formation control without global position information.
Secondly,  the distributed finite-time differentiator is employed  to design  a new distributed finite-time  consensus control algorithm, which can  achieve finite-time  output consensus of  a class of  leader-follower  MAS under  disturbance. Unlike the discontinuous consensus controllers \cite{Cao-Ren-tac-2012,Cao-Ren-tac-2013}, the consensus controller proposed in this paper is continuous, which not only completely suppresses  disturbance, but also avoids chattering.

Notations: For  any  vector  $ {\bf x}=[x_1, x_2,..., x_n]^T\in R^n$, we give some notations.  \\
(1) $\lfloor x_i\rceil^a={\rm sgn}(x_i)\vert x_i\vert^a$, $a\ge0$.
Especially, when $a=0$, define $\lfloor x_i\rceil^0={\rm sgn}(x_i)$.\\
(2)  ${\rm diag}({\bf x})\in R^{n\times n}$ indicates the diagonal matrix with the diagonal element of vector ${\bf x}$.\\
(3) $\lfloor {\bf x}\rceil^a=\Big[\lfloor x_1\rceil^a, \lfloor x_2\rceil^a,..., \lfloor x_n\rceil^a\Big]^T\in R^n$, $a\ge0$.\\
(4)  If matrix $Q=Q^T\in R^n$ is  positive definite,  it is recorded as $Q > 0$,
and the eigenvalues of matrix $Q$  are sorted by size, where the maximum and minimum values are recorded as $\lambda_n(Q)$ and $\lambda_1(Q)$ respectively.\\
(5) Denote ${\bf 1} = [1, . . . , 1]^T$ and ${\bf 0} = [0, . . . , 0]^T $ with appropriate dimension.

\vspace{1.5ex}

\section{ Necessary preparation   }

\subsection{Graph theory}

The  directed graph is often  used to describe  the communication topology  of MAS. $\Psi=\{V,E,A\}$ represents the connectivity among  agents.  $V=\{v_i,i=1,\cdots,n\}$ is the set of vertices, $A=[a_{ij}]\in R^{n\times n}$ is the weighted adjacency matrix and $E\subseteq V\times V$ is the set of edges. Define $\Gamma=\{1,\cdots,n\}$ as node indexes. If $(v_j,v_i)\in E$, then $a_{ij}>0$ and agent $v_j$ is a neighbor agent of agent $v_i$; otherwise, $a_{ij}=0$. The set of all neighboring agents of agent $v_i$ is represented by $N_i=\{j:(v_j,v_i)\in E\}$. The output degree of $v_i$ is defined as: ${\rm deg}_{ out}(v_i)=d_i=\sum_{j=1}^{n}a_{ij}=\sum_{j\in N_i}a_{ij}$.  $D={\rm diag}\{ d_1,\cdots,d_n\}$ is called as the degree matrix. Then $L=D-A$ is called as the Laplacian matrix. The path from $v_i$ to $v_j$ in the graph $\Psi$ is a sequence of different vertices, which starts with $v_i$ and ends with $v_j$ and each step is included in the set $E$. A directed graph $\Psi$ is said to be {\it strongly connected} if there is a path from $v_i $ to $v_j $ between each pair of distinct vertices $v_i $, $v_j $. In addition, if $A^T=A$, then $\Psi(A)$ is said to be an undirected graph. For every different vertices $v_i$ and $v_j$, there is a path from $v_i$ to $v_j$, then $\Psi$ is said to be {\it connected}. If there is a leader, the connectivity between the leader and each follower agent is represented by vector ${\bf b}=[b_1, b_2,..., b_n]^T\in R^n$. If agent $i$ can get the leader's information, then $b_i>0$, otherwise, $b_i=0$. Besides, define $B={\rm diag}({\bf b})$.


\subsection{Some useful lemmas }

\begin{lemma}\label{lf} \cite{bhat_siam_2000} Consider  the following system
\begin{equation}\label{e1}
\dot x=f(x),\quad f(0)=0,\quad x\in R^n, \end{equation} where $f(\cdot):{R}^n\rightarrow {R}^n$  is a continuous function. Suppose there exist a positive definite continuous function $V(x):
U\rightarrow R$, real numbers $c>0$ and $\alpha \in (0,1)$, and an open neighborhood $U_0 \subset U$ of the origin such that $\dot{V}(x)+c(V(x))^\alpha\leq0, x\in U_0\backslash\{0\}.$ Then
$V(x)$ approaches  $0$ in a finite time. In addition, the finite settling time $T $ satisfies that $T\leq \frac{V(x(0))^{1-\alpha}}{c(1-\alpha)}$.
\end{lemma}


%
%
%

\begin{lemma}\label{p+} \cite{qian_2001} Let $c,d>0$. For any $\gamma>0$, the following
inequality holds for $\forall x, y\in {R}$:
 $$|x|^c|y|^d\leq\frac{c}{c+d}\gamma|x|^{c+d}+\frac{d}{c+d}\gamma^{-c/d}|y|^{c+d}.$$
\end{lemma}

\begin{lemma}
\label{pn} \cite{hardy}   For any $x_i\in R,i=1,\cdots,n,$ and  a
real number $p\in (0,1]$,
$$\big(\sum\limits_{i=1}^{n}|x_i|\big)^p\le\sum\limits_{i=1}^{n}|x_i|^p\le n^{1-p}\big(\sum\limits_{i=1}^{n}|x_i|\big)^p.$$
\end{lemma}

\begin{lemma}
\label{pn>1} \cite{hardy}
For any $x_i\in R,i=1,\cdots,n,$ and  a real number $p\ge1$,
$$n^{1-p}\big(\sum\limits_{i=1}^{n}|x_i|\big)^p\le\sum\limits_{i=1}^{n}|x_i|^p\le \big(\sum\limits_{i=1}^{n}|x_i|\big)^p.$$
\end{lemma}

\begin{lemma}\label{pro}\cite{Wang_2010}
If the  directed graph $\Psi$(A) is strongly connected, then there is a column vector ${\bf w}=[w_1, w_2,...,w_n]^T\in R^n$ with all positive elements such that ${\bf w}^TL(A)={\bf 0}^T$.
 Specifically, set $||{\bf w}||_\infty =1$.
In addition,   for a nonnegative vector ${\bf b} \in R^n$, if there exists $ b_i>0$, then
the matrix $G=\frac{1}{2}\big({\rm diag}({\bf w})L(A)+L(A)^T{\rm diag}({\bf w})\big)+{\rm diag}({\bf w}){\rm diag}({\bf b})$ is positive definite.
 Specifically,  ${\bf w} ={\bf 1}$,  if the communication topology is undirected and connected.
\end{lemma}

\vspace{1.5ex}

\section{ Motivations }

For better explanation, we first give the definitions of centralized finite-time differentiator and distributed finite-time differentiator.

\begin{definition}{\bf (Centralized finite-time differentiator)} \label{Centralized-differentiator} \cite{Levant_1998,Levant_2003,Moreno-Fridman_tac-2011,Moreno-Fridman_auto-2013} The differentiator means that for a real-time measurable signal $f(t)$, design an algorithm to estimate $\dot f(t)$ in a finite time under the  condition $\vert\ddot f(t)\vert\le l$, where $l$ is a known positive constant.
\end{definition}
\begin{definition}{\bf (Distributed finite-time differentiator)} \label{Distributed-differentiator}  The differentiator is a distributed sensor network composed of multiple agents. As long as some of agents (at least one agent) can directly measure the signal $f(t)$, then all agents can obtain exact estimates of $f(t)$ and $\dot f(t)$ in a finite time  under condition $\vert\ddot f(t)\vert\le l$, where $l$ is a known positive constant.
\end{definition}

Similarity and difference of  two kind of differentiators are as follows.

{\bf Similarity.} Only signal $f(t)$ is available  under the condition $\vert\ddot f(t)\vert\le l$, while  $\dot f(t)$ and $\ddot f(t)$ are not available.

{\bf Difference.} The centralized finite-time differentiator means that each agent can obtain  the signal $f(t)$, while some of agents (at least one agent) can get the signal $f(t)$ for  the case of  distributed finite-time differentiator.

Centralized finite-time  differentiator
is  generally implemented by second-order sliding mode algorithms  or higher-order sliding mode algorithms, which can be used for state observer design, disturbance observation, and output feedback control
\cite{Levant_1998,Levant_2003,Moreno-Fridman_tac-2011,Moreno-Fridman_auto-2013,Moreno-Fridman-C-2011}. However, centralized finite-time  differentiator is
 not suitable for the distributed  case, while the main aim of  this paper is to solve  this  problem. Besides, for the leader-follower MAS,  in some practice, only the relative position information can be obtained rather than  the absolute global position \cite{Liu-Jiang-auto-2013}. For example, for a group mobile  robots, based on the vision sensor, the relative position information can be easily got. Motivated by above analysis, the  distributed finite-time differentiator via relative position information  is also  proposed.

%
%

\section{ Distributed finite-time differentiator}

\subsection{Problem statement }

Assume that the leader's and i-th agent's positions are  $f(t)$ and $x_i(t)$,  respectively. The main aim of this paper is to
\begin{itemize}
  \item design a distributed finite-time differentiator via relative position information (DFD-R),
  \item design a distributed finite-time differentiator via absolute  position information (DFD-A),
  \item extend DFD to controller form, which will solve the  finite-time output consensus problem of a class of leader-follower MAS.
\end{itemize}

As that  in \cite{Silm-Tac-2019, Du-Wen-auto-2020, Wen-Yu-TIE-2018},
 the communication network of MAS satisfies the following assumption.
\begin{assumption}\label{assum0}
The communication topology of follower agents is {\it strongly connected} and at least one agent can directly obtain the relative or absolute position
information of leader in real time.
\end{assumption}

\begin{assumption}\label{assum0-ab}
The  acceleration information of  leader agent  is bounded, i.e.,
\begin{align}\label{ddot-f}
\vert\ddot{f}(t)\vert\le l,
\end{align}
where $l$ is a positive constant.
\end{assumption}

\subsection{Design of a distributed finite-time differentiator via relative position information}

The dynamics of $i$-th follower agent is assumed to have the form of
\begin{align}\label{dynamics}
\ddot x_i(t)=u_i(t)+\delta_i(t),~~i\in\Gamma,
\end{align}
where $x_i(t)$ is the position, $u_i(t)$ is the control input, $\delta_i(t)$ is the external disturbance which satisfies the following assumption.

\begin{assumption}\label{assum0-disturbance}
The  external disturbance of  each follower agent  is bounded, i.e.,
\begin{align}\label{disturbance}
\vert\delta_i(t)\vert\le l_1,~~i\in\Gamma,
\end{align}
where $l_1$ is a positive constant.
\end{assumption}

For each follower agent, a DFD-R  is designed as follows
\begin{align}\label{differentiator}
\dot {\hat{p}}_i=&{\hat{q}}_i-k_1\lfloor y_i\rceil^{\frac{1}{2}}, ~~
\dot {\hat{q}}_i=-k_2\lfloor y_i\rceil^0{+u_i},
\end{align}
where
\begin{align}
y_i=\sum\limits_{j\in N_i}a_{ij}\big(\hat{p}_i-\hat{p}_j-(x_i-x_j)\big)+b_i\big(\hat{p}_i-(x_i-f)\big).
\end{align}

\begin{theorem}\label{theo1}
For  MAS under  Assumptions \ref{assum0}-\ref{assum0-disturbance}, if the DFD-R is designed as (\ref{differentiator}) and parameters $k_1, k_2$ are selected as
\begin{align}\label{k1k2}
&k_2\ge\frac{l_2}{\rho},~~~
k_1\ge \Big(\frac{2(\gamma_0+\gamma_1)+1}{\lambda_1(G)}\Big)^{\frac{1}{2}}k_2^{\frac{1}{2}},
\end{align}
where  {$l_2=l+l_1$, }$0<\rho<1$, $\gamma_0=(1+3\gamma_1)/(1-\rho)$, $\gamma_1=(1+\rho)\max\{w_i\}$, 
then each follower agent can estimate the relative position and relative velocity between leader and itself in a finite time,
i.e.,  ${\hat{p}}_i\rightarrow (x_i-f)$, ${\hat{q}}_i\rightarrow (\dot x_i-\dot f)$ in a finite time.
\end{theorem}
 {\bf Proof : }
Define the estimation error $e_i=\hat{p}_i-(x_i-f),~~z_i={\hat{q}}_i-(\dot x_i-\dot f).$
Hence,  the error equation is given  as follows
\begin{align}\label{model}
\dot {\bf e}=&{\bf z}-k_1\lfloor {\bf y}\rceil^{\frac{1}{2}}, ~~~~
\dot {\bf z}=-k_2\lfloor {\bf y}\rceil^{0}+{\bf d},
\end{align}
where ${\bf e}=[e_1, e_2,..., e_n]^T\in R^n$, ${\bf y}=[y_1, y_2,..., y_n]^T\in R^n$, ${\bf z}=[z_1, z_2,..., z_n]^T\in R^n$, ${\bf d}=[d_1, d_2,..., d_n]^T\in R^n,  d_i=\ddot{f}(t)-\delta_i(t)$.

By noticing  that $y_i=\sum\limits_{j\in N_i}a_{ij}(e_i-e_j)+b_ie_i$, then
\begin{align}\label{x-y}
{\bf y}=(L+B){\bf e}.
\end{align}
Letting  ${\bf v}=\frac{{\bf z}}{k_1}, k=\frac{k_2}{k_1}$,  then
\begin{align}\label{model-1}
\dot {\bf e}=&k_1({\bf v}-\lfloor {\bf y}\rceil^{\frac{1}{2}}),~~~
\dot {\bf v}=k(-\lfloor {\bf y}\rceil^{0}+\frac{{\bf d}}{k_2}).
\end{align}
It is easy to know  ${\vert d_i\vert}/{k_2}\le{{l_2}}/{k_2}\le\rho<1$.
The Lyapunov function is constructed as
\begin{align}\label{V}
V=V_1+\gamma_0V_2,
\end{align}
where
\begin{align}\label{V1-V2}
V_1=\sum_{i=1}^{n}w_i\int_{\lfloor v_i\rceil^2}^{y_i}(\lfloor s\rceil^{\frac{1}{2}}-v_i)\mathrm{d}s,
V_2=\frac{1}{3}\sum_{i=1}^{n}\vert v_i\vert^3.
\end{align}

The first step is to obtain the derivative of $V_1$, i.e.,

\begin{align}\label{dot-V1-0}
\dot V_1=\sum_{i=1}^{n}w_i(\lfloor y_i\rceil^{\frac{1}{2}}-v_i)\dot y_i-\sum_{i=1}^{n}w_i(y_i-\lfloor v_i\rceil^2)\dot v_i\nn\\
\le\sum_{i=1}^{n}w_i(\lfloor y_i\rceil^{\frac{1}{2}}-v_i)\dot y_i+k\gamma_1\sum_{i=1}^{n}\vert y_i-\lfloor v_i\rceil^2\vert.
\end{align}

For the  first term, by Lemma \ref{pro}, one has that
\begin{align}\label{dot-V1-1}
&\sum_{i=1}^{n}w_i(\lfloor y_i\rceil^{\frac{1}{2}}-v_i)\dot y_i\nn\\
=&-({\bf v}-\lfloor {\bf y}\rceil^{\frac{1}{2}})^T{\rm diag}({\bf w})\dot {\bf y}\nn\\
=&-k_1({\bf v}-\lfloor {\bf y}\rceil^{\frac{1}{2}})^T{\rm diag}({\bf w})(L+B)({\bf v}-\lfloor {\bf y}\rceil^{\frac{1}{2}})\nn\\
=&-k_1({\bf v}-\lfloor {\bf y}\rceil^{\frac{1}{2}})^TG({\bf v}-\lfloor {\bf y}\rceil^{\frac{1}{2}})\nn\\
\le&-k_1\lambda_1(G)\sum_{i=1}^{n}\vert v_i-\lfloor y_i\rceil^\frac{1}{2}\vert^2.
\end{align}
Applying  Lemma \ref{pn>1} to the second term of inequality (\ref{dot-V1-0}) results in
\begin{align}\label{dot-V1-2}
\vert y_i-\lfloor v_i\rceil^2\vert
\le&\vert v_i\vert^2+\vert v_i-\lfloor y_i\rceil^\frac{1}{2}-v_i\vert^2\nn\\
\le&3\vert v_i\vert^2+2\vert v_i-\lfloor y_i\rceil^\frac{1}{2}\vert^2.
\end{align}
Substituting (\ref{dot-V1-1}) and  (\ref{dot-V1-2}) into  (\ref{dot-V1-0}) leads to
\begin{align}\label{dot-V1-3}
\dot V_1\le&-\Big(k_1\lambda_1(G)-2k\gamma_1\Big)\sum_{i=1}^{n}\vert v_i-\lfloor y_i\rceil^\frac{1}{2}\vert^2\nn\\
&+3k\gamma_1\sum_{i=1}^{n}\vert v_i\vert^2.
\end{align}
The second step is to get  the derivative of $V_2$, i.e.,
\begin{align}\label{dot-V2-0}
\dot V_2&=\sum_{i=1}^{n}\lfloor v_i\rceil^2k(-\lfloor y_i\rceil^0+\frac{d_i}{k_2})\nn\\
&=\sum_{i=1}^{n}\lfloor v_i\rceil^2k(\lfloor v_i\rceil^0-\lfloor y_i\rceil^0-\lfloor v_i\rceil^0+\frac{d_i}{k_2})\nn\\
&\le-k(1-\rho)\sum_{i=1}^{n}\vert v_i\vert^2+k\sum_{i=1}^{n}\vert v_i\vert^2\vert \lfloor y_i\rceil^0-\lfloor v_i\rceil^0\vert.
\end{align}
Next, we will estimate the last term of inequality (\ref{dot-V2-0}) in two cases.
{\bf Case 1:} If  $y_iv_i>0$, then $\lfloor y_i\rceil^0-\lfloor v_i\rceil^0=0$.
{\bf Case 2:} If  $y_iv_i\le0$, then $\vert\lfloor y_i\rceil^0-\lfloor v_i\rceil^0\vert\le2$,
 $\vert v_i\vert\le\vert v_i-\lfloor y_i\rceil^\frac{1}{2}\vert$.
In both cases, the following inequality always holds
\begin{align}\label{A1}
\vert v_i\vert^2\vert \lfloor y_i\rceil^0-\lfloor v_i\rceil^0\vert\le2\vert v_i-\lfloor y_i\rceil^\frac{1}{2}\vert^2.
\end{align}
Substituting (\ref{A1}) into (\ref{dot-V2-0}) leads to
\begin{align}\label{dot-V2-1}
\dot V_2\le-k(1-\rho)\sum_{i=1}^{n}\vert v_i\vert^2+2k\sum_{i=1}^{n}\vert v_i-\lfloor y_i\rceil^\frac{1}{2}\vert^2.
\end{align}
To sum up, we have
\begin{align}\label{dot-V-0}
\dot V\le&-\Big(k_1\lambda_1(G)-2k(\gamma_0+\gamma_1)\Big)\sum_{i=1}^{n}\vert v_i-\lfloor y_i\rceil^\frac{1}{2}\vert^2\nn\\
&-k\sum_{i=1}^{n}\vert v_i\vert^2.
\end{align}
Using the gain condition  (\ref{k1k2})  and $k=\frac{k_2}{k_1}$  leads to
\begin{align}\label{dot-V-1}
\dot V\le&-k\Big(\sum_{i=1}^{n}\vert v_i-\lfloor y_i\rceil^{\frac{1}{2}}\vert^2+\sum_{i=1}^{n}|v_i|^2\Big).
\end{align}
On the other hand,  one has
\begin{align}\label{V1}
V_1
\le&\sum_{i=1}^{n}w_i\vert y_i-\lfloor v_i\rceil^2\vert\vert v_i-\lfloor y_i\rceil^{\frac{1}{2}}\vert\nn\\
\le&\frac{\gamma_1}{1+\rho}\sum_{i=1}^{n}\vert y_i-\lfloor v_i\rceil^2\vert\vert v_i-\lfloor y_i\rceil^{\frac{1}{2}}\vert.
\end{align}
By  inequality  (\ref{dot-V1-2}) and Lemma \ref{p+}, one obtains
\begin{align}\label{B1}
&\vert y_i-\lfloor v_i\rceil^2\vert\vert v_i-\lfloor y_i\rceil^{\frac{1}{2}}\vert\nn\\
\le&3\vert v_i\vert^2\vert v_i-\lfloor y_i\rceil^{\frac{1}{2}}\vert+2\vert v_i-\lfloor y_i\rceil^\frac{1}{2}\vert^3\nn\\
\le&2\vert v_i\vert^3+3\vert v_i-\lfloor y_i\rceil^\frac{1}{2}\vert^3.
\end{align}
Substituting this inequality into  (\ref{V1}) leads to
\begin{align}\label{V-1}
V\le\gamma_2\Big(\sum_{i=1}^{n}\vert v_i-\lfloor y_i\rceil^\frac{1}{2}\vert^3+\sum_{i=1}^{n}\vert v_i\vert^3\Big),
\end{align}
where $\gamma_2=\frac{2\gamma_1}{1+\rho}+\frac{\gamma_0}{3}$. Furthermore, it follows from  Lemma \ref{pn} that
\begin{align}\label{V-2}
\Big(\frac{V}{\gamma_2}\Big)^\frac{2}{3}\le\sum_{i=1}^{n}\vert v_i-\lfloor y_i\rceil^\frac{1}{2}\vert^2+\sum_{i=1}^{n}\vert v_i\vert^2.
\end{align}
As a result, substituting (\ref{V-2}) into (\ref{dot-V-1}) leads to
\begin{align}\label{dot-V-2}
\dot V\le-k\gamma_2^{-\frac{2}{3}}V^{\frac{2}{3}},
\end{align}
which implies that $V$ will converge to 0 in a finite time  and the setting time  $T $ satisfies that $T\leq 3\gamma_2^{\frac{2}{3}}V(0)/k$. In other words,  it means that  ${\bf v}={\bf y}={\bf 0}$, which implies that ${\bf z}={\bf 0}$.   Furthermore, from Lemma 2.5, it can be seen that $G=\frac{1}{2}\big({\rm diag}({\bf w})L(A)+L(A)^T{\rm diag}({\bf w})\big)+{\rm diag}({\bf w})B$ is positive definite, and thus it is easy to obtain that $L+B$ is a nonsingular matrix. Therefore, one has that ${\bf e}=(L+B)^{-1}{\bf y}={\bf 0}$.
$\blacksquare$

\begin{remark}
If the communication topology is undirected and connected, according to Lemma 2.5, we have $w ={\bf 1}$, i.e., $ w_i=1$. This means that $\gamma_1=1+\rho$, $G=L(A)+{\rm diag}({\bf b})$, and the Lyapunov function (\ref{V}) is also simplified, which makes the
 subsequent proofs simpler. To avoid repetition, the proof is  omitted.
\end{remark}

\begin{remark}
In some situations, such as without GPS and other global measuring equipment, the absolute global position and absolute global velocity cannot be obtained. Our proposed algorithm only needs the relative position information, which is more suitable for some practical situations\cite{Liu-Jiang-auto-2013,Lv-Wen-auto-2020,du-wen-yu-auto-2015,Cao-Ren-SCL-2010}.
Besides, many formation tracking/flying scenarios can be divided into two parts: distributed state estimation and desired state tracking by  only using relative position information, which  has important theoretical significance and practical value.
\end{remark}

\subsection{Design of distributed finite-time differentiator via absolute  position information}



\begin{theorem}\label{theo2}
For  MAS under  Assumptions \ref{assum0} and \ref{assum0-ab}, if the DFD-A is designed  as
\begin{align}
\dot {\hat{p}}_i=&{\hat{q}}_i-k_1\lfloor y_i\rceil^{\frac{1}{2}}, ~~
\dot {\hat{q}}_i=-k_2\lfloor y_i\rceil^0,\nn\\
y_i=&\sum\limits_{j\in N_i}a_{ij}(\hat{p}_i-\hat{p}_j)+b_i(\hat{p}_i-f),~~~i\in\Gamma, \label{differentiator-abso}
\end{align}
where parameters $k_1, k_2$ are selected as (\ref{k1k2}),
then the distributed finite-time differential estimation  is realized, i.e.,
 ${\hat{p}}_i\rightarrow f$, ${\hat{q}}_i\rightarrow \dot f$ in a finite time.
\end{theorem}
 {\bf Proof : }
For each agent, defining  $e_{i}={\hat{p}}_i-f$, ${z_i}={\hat{q}}_i-\dot f$, $d_i=-\ddot f$, then
one has
\begin{align}
\dot e_i=&z_i-k_1\lfloor y_i\rceil^{1/2},~~~
\dot z_i=-k_2\lfloor y_i\rceil^{0}+d_i,\nn\\
y_i=&\sum\limits_{j=1}^na_{ij}(e_i-e_j)+b_ie_i, ~i\in\Gamma,\end{align}
or in the form of vector
\begin{align}\label{model-1-2}
\dot {\bf e}=&{\bf z}-k_1\lfloor {\bf y}\rceil^{1/2},~~~
\dot {\bf z}=-k_2\lfloor {\bf y}\rceil^{0}+{\bf d}.
\end{align}
According to Assumption  \ref{assum0-ab}, it is evident that $\vert d_i\vert\le l$.   By using a  same proof  as  that in Theorem \ref{theo1}, it can be proved that  the system
(\ref{model-1-2}) is finite-time stable. $\blacksquare$

\begin{remark}
Compared to the first-order observer and second-order observer proposed by \cite{Cao-Ren-SCL-2010}, the distributed finite-time differentiator presented in this paper has three  differences. Firstly, the  proposed method in this paper  only assumes that $\ddot f$ is bounded, without any additional assumptions on the own and neighbors' velocity. Secondly, the estimation result of $\hat q_i$ is always  continuous and converges to $\dot f$ in a finite time. Thirdly, for a first-order multi-agent system, as demonstrated in Theorem 5.1, the proposed distributed finite-time differentiator algorithm can be used to design a continuous finite-time consensus controller. However, if the first-order observer of \cite{Cao-Ren-SCL-2010} is used, only a discontinuous finite-time consensus controller can be designed to suppress the disturbances.
\end{remark}

\begin{remark}
The main difference between the two DFDs lies in their usage conditions or scenarios. DFD-A utilizes the leader's global absolute position information, enabling all follower agents to obtain the leader's global absolute position information and global absolute velocity information. In contrast, DFD-R utilizes the relative position information with the leader, enabling all follower agents to obtain the relative position and relative velocity information with the leader. However, the connection between the two DFDs lies in two aspects. Firstly, they are both distributed differential estimation algorithms, i.e., distributed finite-time differentiator. Secondly, they share the same mathematical essence, i.e., equation (\ref{model}).
\end{remark}

\vspace{1.5ex}

\section{\bf Design of   distributed finite-time consensus controller  for  leader-follower  MAS  }

Inspired by the duality principle, we will show that how to extend the DFD to a new distributed finite-time  consensus control algorithm such that all agents' output can achieve consensus in a  finite time. Without loss of generality, the dynamics of follower agent $i$ is given as follow
\begin{align}\label{follower-1}
&\dot x_i(t)=f_i(t, x_i)+g_i(t, x_i)u_i, \nn\\
&\ s_i=s_i(t, x_i),
\end{align}
where $n_i$ is the order of system, $f_i(t, x_i), g_i(t, x_i)\in R^{n_i}$ are smooth vector functions, $x_i\in R^{n_i}$ is the state vector,
 $u_i\in R$ is the control input, and $ s_i\in R$ is the output.
Assume that  the relative degree of output is  one with regard to  control input, i.e.,
\begin{align}\label{follower-1-2}
\dot s_i=a_i(t, x_i)+b_i(t, x_i)u_i,
\end{align}
where $a_i(t, x_i)$ is an unknown smooth function including possible uncertainties and  external disturbance, etc.,    $b_i(t, x_i)>0$ is a known function.

The dynamics of leader agent is as follow
\begin{align}\label{leader-1}
\dot x_0(t)=f_0(t, x_0), \ s_0=s_0(t, x_0),\
\dot s_0=a_0(t, x_0),
\end{align}
where $n_0$ is the order of system, $f_0(t, x_0)\in R^{n_0}$ is a smooth vector function, $x_0\in R^{n_0}$ is the state vector, $ s_0\in R$ is the output, and $a_0(t, x_0)$ is an unknown smooth function.

\begin{remark}
Note that  for any different agent $i$ and agent $j$,
 the functions $f_i$, $g_i$, $s_i$,  $a_i$, $b_i$ and system's order  $n_i$ can be different from $f_j$, $g_j$, $s_j$,  $a_j$, $b_j$ and $n_j$, respectively.
 It means that the dynamics of each agent can be completely different, i.e.,  heterogeneous.
\end{remark}


\begin{assumption}\label{assum2}
For $\forall i\in\Gamma$, $\vert \dot a_i(t, x_i)-\dot a_0(t, x_0)\vert\le l$, $l$ is a positive constant.
\end{assumption}

%
\begin{theorem}\label{theo3}
For  the leader-follower  MAS (\ref{follower-1})-(\ref{leader-1}) under  Assumptions \ref{assum0} and  \ref{assum2}, if the distributed controller  is designed as
\begin{align}\label{u-1}
u_i=&\frac{1}{b_i(t, x_i)}\Big(v_i-k_1\lfloor y_i\rceil^{1/2}\Big),~~
\dot v_i=-k_2\lfloor y_i\rceil^0,\nn\\
y_i&=\sum\limits_{j\in N_i}a_{ij}(s_i-s_j)+b_i(s_i-s_0), ~~~i\in\Gamma,\end{align}
where the  parameters $k_1, k_2$ are selected as (\ref{k1k2}), then
  the output  of all followers' agents can track the leader's output  in a finite time,
i.e.,  $s_i\rightarrow s_0$ in a finite time.
\end{theorem}
 {\bf Proof : }
For each agent, define  $e_{i}=s_i-s_0$, $z_i=v_i+a_i-a_0$, $d_i=\dot a_i-\dot a_0$,  then
\begin{align}
\dot e_i=&z_i-k_1\lfloor y_i\rceil^{1/2},~~~
\dot z_i=-k_2\lfloor y_i\rceil^{0}+d_i, ~~i\in\Gamma,\nn\\
y_i
=&\sum\limits_{j=1}^na_{ij}(e_i-e_j)+b_ie_i,
\end{align}
or in the vector form
\begin{align}\label{model-2-1-1}
\dot {\bf e}=&{\bf z}-k_1\lfloor {\bf y}\rceil^{1/2},~~~
\dot {\bf z}=-k_2\lfloor {\bf y}\rceil^{0}+{\bf d}.
\end{align}
 According to Assumption \ref{assum2}, it is evident that $\vert d_i\vert\le l$.
As a sequel, the following proof can be achieved
by using a  same proof as  (\ref{model}) in
 Theorem \ref{theo1} and is omitted here. $\blacksquare$

\begin{remark} Actually, for the consensus tracking problem   of  MAS (\ref{follower-1})-(\ref{leader-1}), based on the
 variable structure control method, the finite-time consensus can be also achieved \cite{Cao-Ren-tac-2012}.
Inspired by but different from discontinuous consensus controllers\cite{Cao-Ren-tac-2012,Cao-Ren-tac-2013}, the consensus controller proposed in this paper is continuous, which not only completely suppresses  disturbance, but also avoids chattering.
\end{remark}

\vspace{1.5ex}

\section{\bf Numerical examples and simulations}

One  typical communication topology is shown in Fig. 1.

\begin{figure}[h]
\begin{center}
\includegraphics[height=3.2cm]{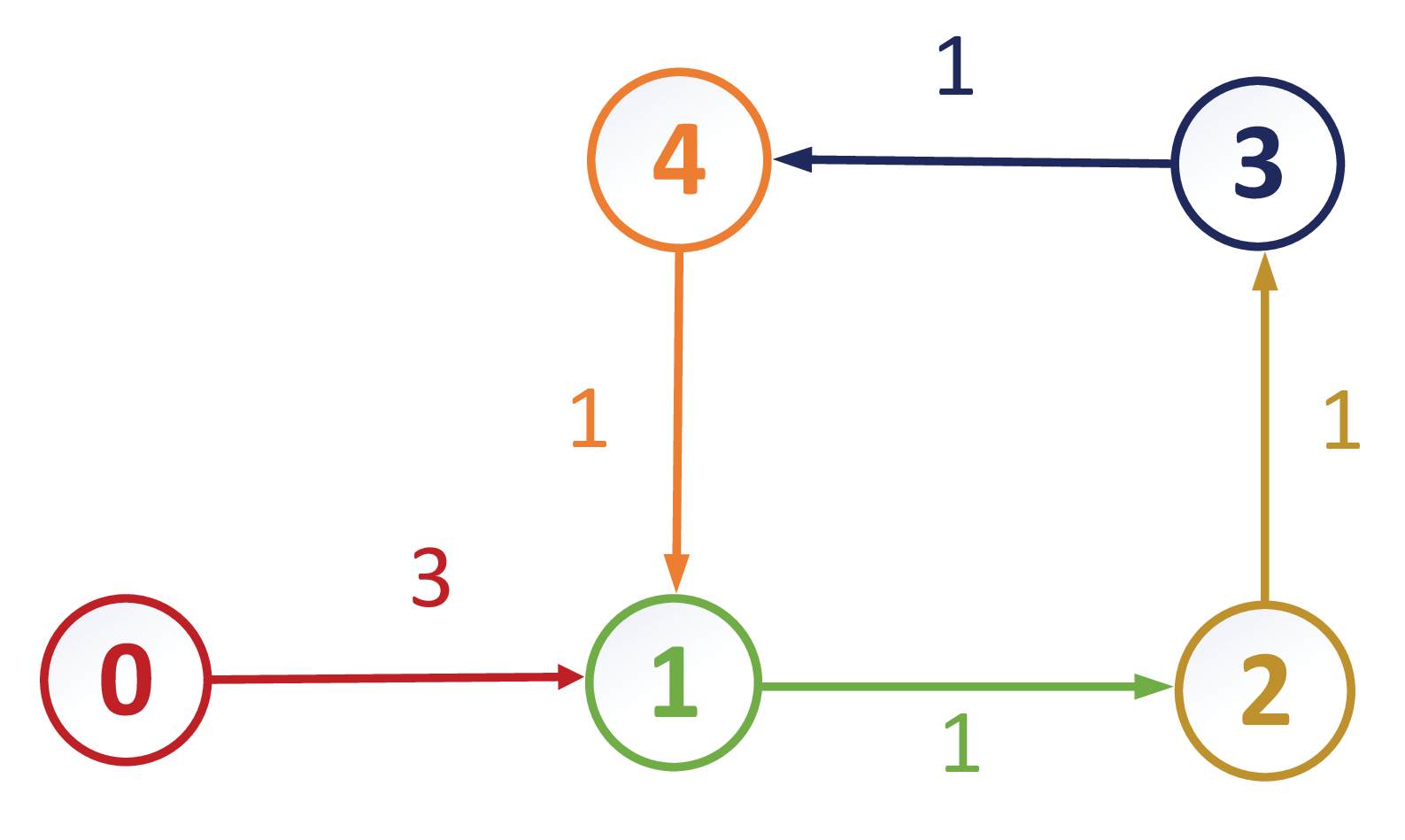}
\caption{Communication topology A among agents.}
\end{center}
\label{Fig1-1}
\end{figure}

\subsection{\bf Distributed finite-time differentiator via relative position information}

\begin{figure}
\centering \epsfxsize=3.6 in \epsfysize=2.6 in
\epsfclipon \epsffile{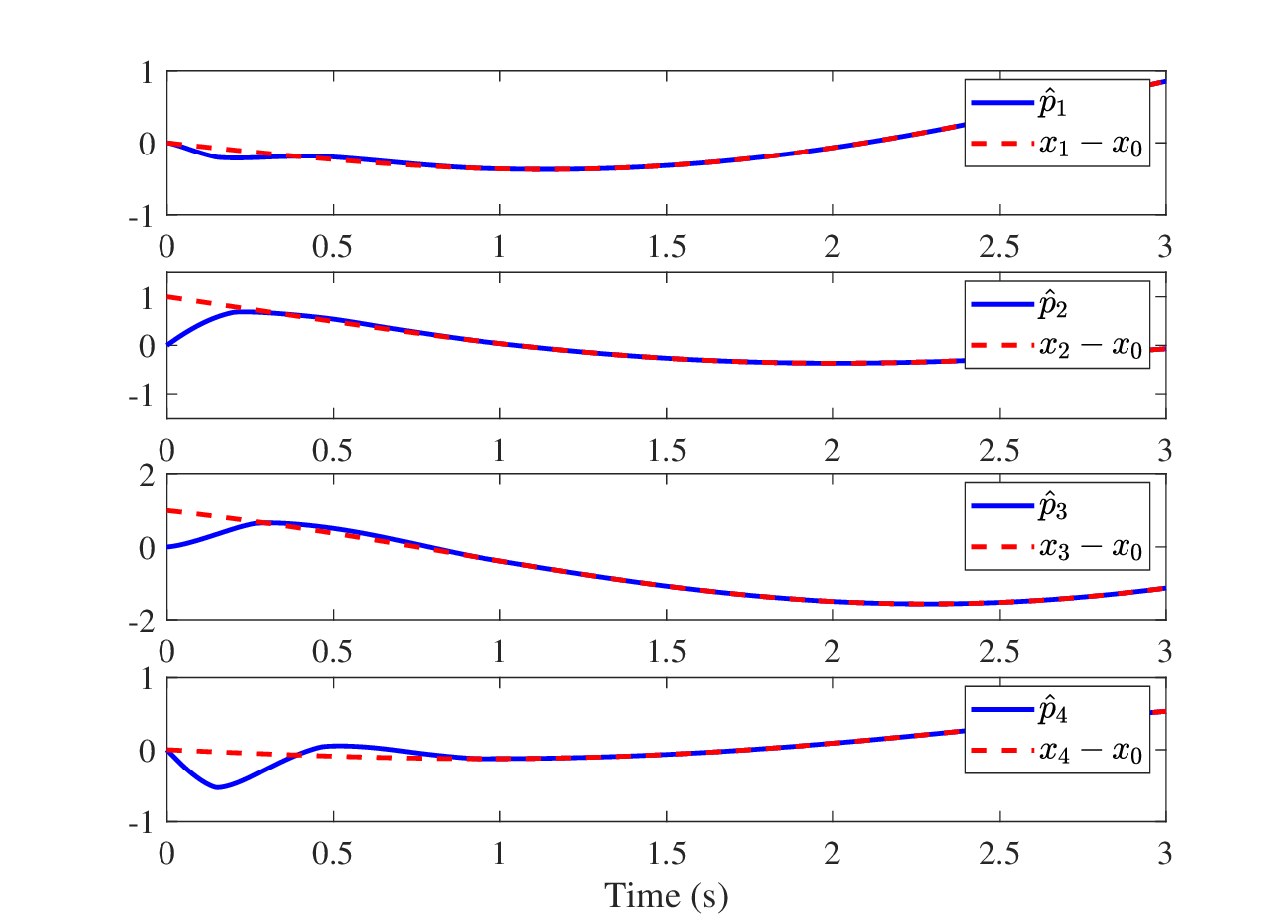}
\caption{ The response curves of relative position estimation under communication topology A.}
\label{Fig2-1}
\end{figure}

\begin{figure}
\centering \epsfxsize=3.6 in \epsfysize=2.6 in
\epsfclipon \epsffile{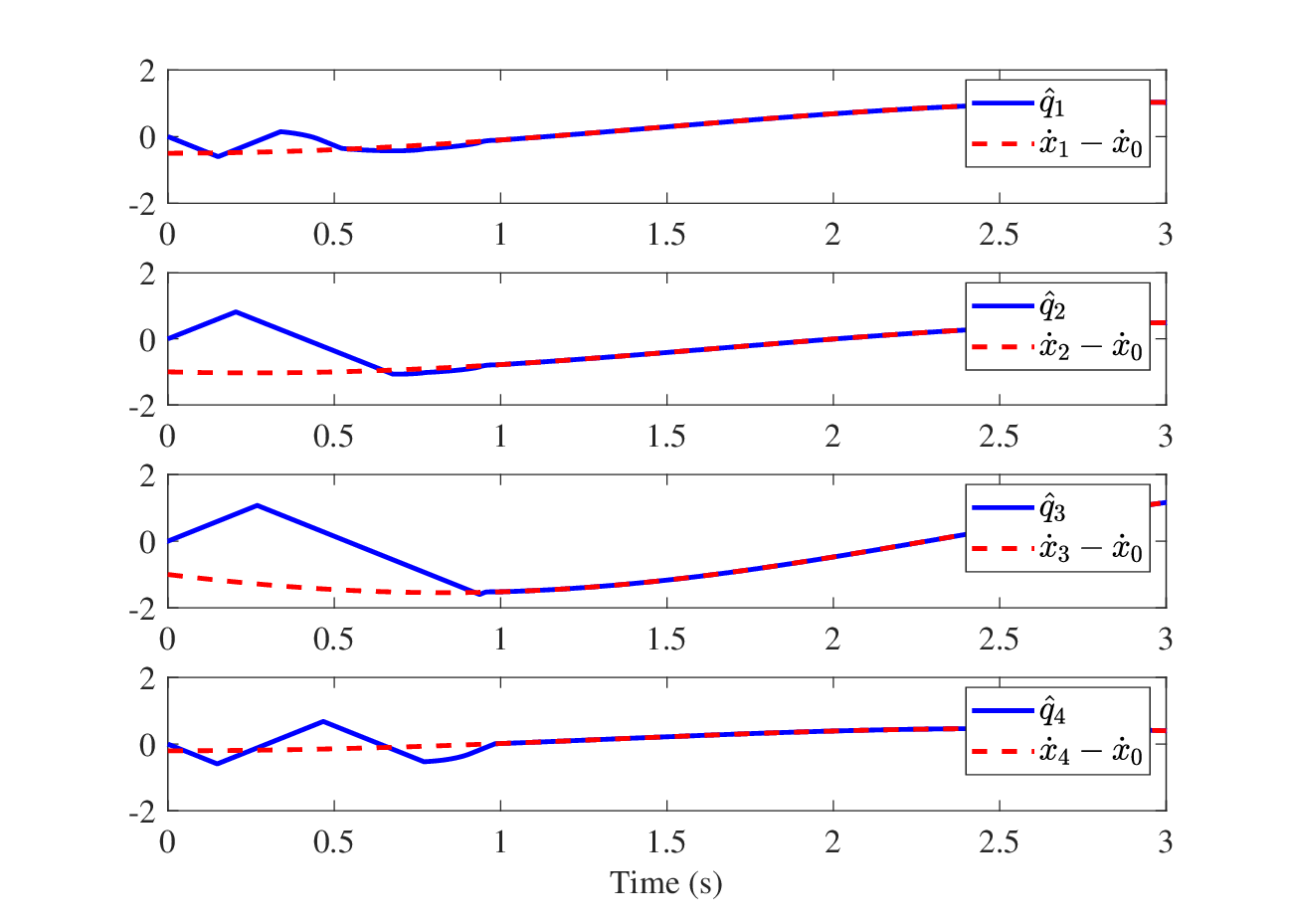}
\caption{ The response curves of relative velocity estimation under communication topology A.}
\label{Fig2-2}
\end{figure}

In the simulation, we set $x_0(t)=\sin(t)$,  $ \delta_1(t)=-0.25\sin(0.5t),$ $ \delta_2(t)=-0.25\cos(0.5t),$ $ \delta_3(t)=-1.21\cos(1.1t),$ $ \delta_4(t)=-0.64\sin(0.8t)$,
 $u_i=0, \forall i\in\{1,2,3,4\}$. The initial values of four followers are set as $x_1(0)=0, x_2(0)=1, x_3(0)=1, x_4(0)=0.$
The initial values of distributed finite-time differentiator (\ref{differentiator}) are set as: $\hat p_i(0)=\hat q_i(0)=0, \forall i\in\{1,2,3,4\}$. The gains of differentiator are selected as: $k_1=5, k_2=4$. The response curves of relative position estimation and relative velocity estimation under  communication topology A  are shown in Fig. \ref{Fig2-1} and Fig. \ref{Fig2-2}, respectively. It can be seen from the figures that each follower agent can estimate the relative position and relative velocity between itself and the leader in a finite time, which verifies the effectiveness of DFD-R.

\subsection{\bf Distributed finite-time differentiator via absolute position information  }

\begin{figure}
\centering \epsfxsize=3.6 in \epsfysize=2.5 in
\epsfclipon \epsffile{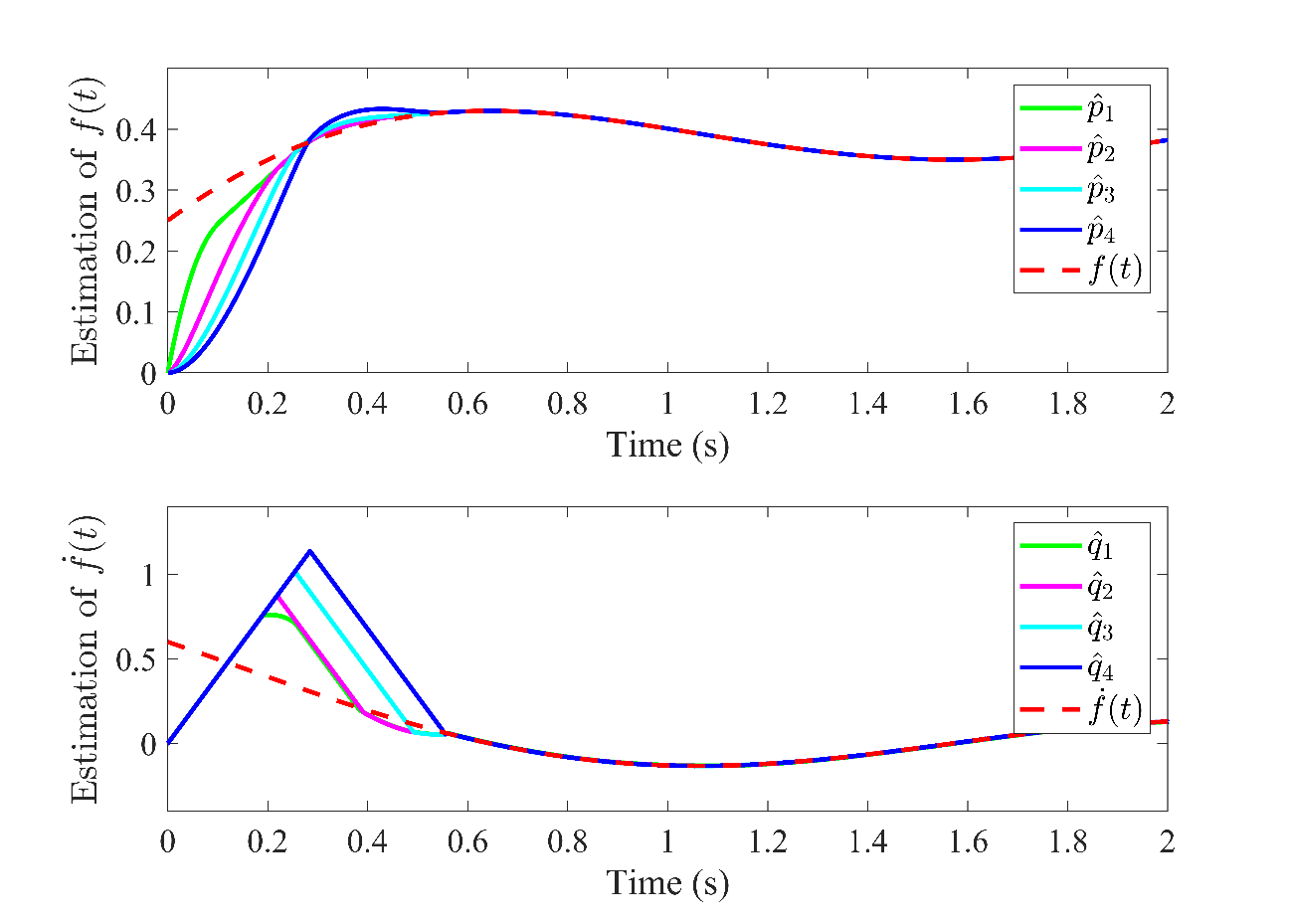}
\caption{ The response curves  of absolute position estimation and absolute velocity estimation under communication topology A.}
\label{Fig3-1}
\end{figure}

The signal to be observed is: $f=0.6\sin(t)+0.25\cos(2t)$, thus $\dot f=0.6\cos(t)-0.5\sin(2t)$ and $\vert\ddot f\vert\le 1.6<l=3$ under a conservative estimate. The initial values of distributed finite-time differentiator (\ref{differentiator-abso}) are set as: $\hat p_i(0)=\hat q_i(0)=0, \forall i\in\{1,2,3,4\}$. The gains of differentiator are selected as: $k_1=5, k_2=4$. The response curves of absolute position estimation and absolute velocity estimation under communication topology A are shown in Fig. \ref{Fig3-1}.
It can be seen from Fig. \ref{Fig3-1} that each follower agent can estimate the absolute position and absolute velocity of leader in a finite time, which verifies the effectiveness of DFD-A.

\subsection{\bf Distributed finite-time  consensus controller  }

\begin{figure}
\centering \epsfxsize=3.7 in \epsfysize=2.6 in
\epsfclipon \epsffile{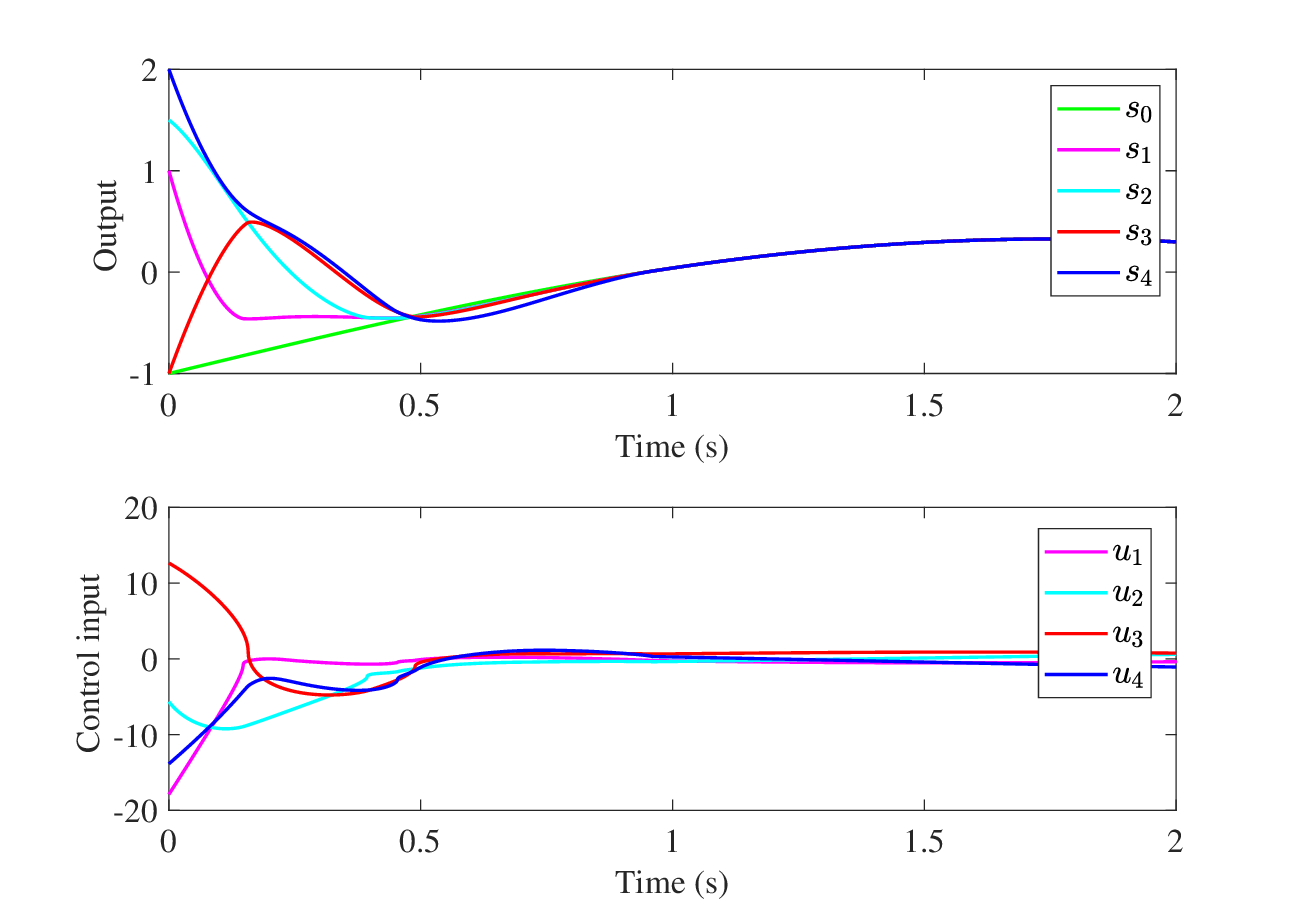}
\caption{ The response curves of MAS' output and control input under communication topology A.}
\label{Fig4-1}
\end{figure}

On the basis of  works  \cite{Cao-Ren-tac-2012,Cao-Ren-auto-2014}, we consider the following leader-follower MAS: $\dot s_0(t)=a_0(t), \dot s_i(t)=a_i(t)+u_i,$
where $s_0$ and $s_i$ are the output of leader agent and i-th follower agent respectively, $a_0$ and $a_i$ are unknown functions with bounded change rate, $u_i$ is the control input of  i-th follower agent.
In this simulation, we set $a_0(t)=\cos (t)+0.2\cos (0.2t), a_1(t)=\sin(1.5t), a_2(t)=2\cos(t), a_3(t)=\cos(1.5t), a_4(t)=\sin(0.5t)$. Then, $\vert \dot a_i(t)-\dot a_0(t)\vert\le 3.1$.
For controller (\ref{u-1}), $b_i=1$ and we set $k_1=8, k_2=6$,  $[v_1(0), v_2(0), v_3(0), v_4(0)]=[0, 0, 0, 0]$. The initial values of the system are set as $[s_0(0), s_1(0), s_2(0), s_3(0), s_4(0)]=[-1, 1, 1.5, -1, 2]$. The response curves of MAS' output and control input under communication topology A are shown in Fig. \ref{Fig4-1}. Note that the controller  is  continuous which   is chattering-free
and  is also  an advantage by comparing  with the  discontinuous controller.

\vspace{1.5ex}

\section{\bf Conclusion }

In this paper, distributed finite-time differentiator (DFD) has been proposed by using relative or absolute position information, and its finite-time stability has been proved by skillfully constructing Lyapunov function. The output consensus of a class of  leader-follower MAS has been achieved  by extending DFD. In the future, we will try to extend the DFD to  higher-order case, and apply the algorithm to  formation coordination control by only using relative position information.


\end{document}